\documentclass[iop]{emulateapj}

\newcommand{\eqref}[1]{(\ref{#1})}

\def\lesssim{\mathrel{\hbox{\rlap{\hbox{\lower4pt\hbox{$\sim$}}}\hbox{$<$}}}}
\def\gtrsim{\mathrel{\hbox{\rlap{\hbox{\lower4pt\hbox{$\sim$}}}\hbox{$>$}}}}

\shorttitle{TWHya}
\shortauthors{Akiyama et al.}

\usepackage{color}
\usepackage{ulem}
\begin{document}


\title{DISCOVERY OF A DISK GAP CANDIDATE AT 20 AU IN TW HYDRAE}


\author{E. AKIYAMA\altaffilmark{1}, T. MUTO\altaffilmark{2}, N. KUSAKABE\altaffilmark{1}, A. KATAOKA\altaffilmark{3}, J. HASHIMOTO\altaffilmark{4}, T. TSUKAGOSHI\altaffilmark{5}, J. KWON\altaffilmark{6}, T. KUDO\altaffilmark{7}, R. KANDORI\altaffilmark{1}, C. A. GRADY\altaffilmark{8}, M. TAKAMI\altaffilmark{9}, M. JANSON\altaffilmark{10}, M. KUZUHARA\altaffilmark{3}, T. HENNING\altaffilmark{11}, M. L. SITKO\altaffilmark{12,13}, J. C. CARSON\altaffilmark{11,14}, S. MAYAMA\altaffilmark{15}, T. CURRIE\altaffilmark{7}, C. THALMANN\altaffilmark{16}, J. WISNIEWSKI\altaffilmark{4}, M. MOMOSE\altaffilmark{5}, N. OHASHI\altaffilmark{7}, L. ABE\altaffilmark{17}, W. BRANDNER\altaffilmark{11}, T. D. BRANDT\altaffilmark{18}, S. EGNER\altaffilmark{7}, M. FELDT\altaffilmark{11}, M. GOTO\altaffilmark{19}, O. GUYON\altaffilmark{7}, Y. HAYANO\altaffilmark{7}, M. HAYASHI\altaffilmark{1}, S. HAYASHI\altaffilmark{7}, K. W. HODAPP\altaffilmark{20}, M. ISHI\altaffilmark{1}, M. IYE\altaffilmark{1}, G. R. KNAPP\altaffilmark{18}, T. MATSUO\altaffilmark{21}, M. W. McELWAIN\altaffilmark{22}, S. MIYAMA\altaffilmark{23}, J. -I. MORINO\altaffilmark{1}, A. MORO-MARTIN\altaffilmark{24,25}, T. NISHIMURA\altaffilmark{7}, T. -S. PYO\altaffilmark{7}, G. SERABYN\altaffilmark{26}, T. SUENAGA\altaffilmark{1,27}, H. SUTO\altaffilmark{1}, R. SUZUKI\altaffilmark{1}, Y. H. TAKAHASHI\altaffilmark{1,6}, N. TAKATO\altaffilmark{7}, H. TERADA\altaffilmark{7}, D. TOMONO\altaffilmark{7}, E. L. TURNER\altaffilmark{18,28}, M. WATANABE\altaffilmark{29}, T. YAMADA\altaffilmark{30}, H. TAKAMI\altaffilmark{1}, T. USUDA\altaffilmark{1}, M. TAMURA\altaffilmark{1,6,31}}

\affil{\altaffilmark{1}National Astronomical Observatory of Japan, 2-21-1, Osawa, Mitaka, Tokyo, 181-8588, Japan; eiji.akiyama@nao.ac.jp \\
\altaffilmark{2}Division of Liberal Arts, Kogakuin University, 1-24-2, Nishi-Shinjuku, Shinjuku-ku, Tokyo, 163-8677, Japan \\
\altaffilmark{3}Department of Earth and Planetary Sciences, Tokyo Institute of Technology, 2-12-1, Ookayama, Meguro-ku, Tokyo 152-8551, Japan \\
\altaffilmark{4}Department of Physics and Astronomy, The University of Oklahoma, 440 W. Brooks St. Norman, OK, 73019, USA \\
\altaffilmark{5}College of Science, Ibaraki University, 2-1-1, Bunkyo, Mito, Ibaraki, 310-8512, Japan \\
\altaffilmark{6}Department of Astronomy, Graduate School of Science, The University of Tokyo, 7-3-1, Hongo, Bunkyo-ku, Tokyo, 113-0033, Japan \\
\altaffilmark{7}Subaru Telescope, National Astronomical Observatory of Japan, 650, North A'ohoku Place, Hilo, HI, 96720, USA \\
\altaffilmark{8}Eureka Scientific, 2452 Delmer, Suite 100, Oakland, CA, 96002, USA \\
\altaffilmark{9}Institute of Astronomy and Astrophysics, Academia Sinica, P.O. Box 23-141, Taipei, 10617, Taiwan \\
\altaffilmark{10}Department of Astronomy, Stockholm University, AlbaNova University Center, Stockholm, 106 91, Sweden \\
\altaffilmark{11}Max Planck Institute for Astronomy, K\"{o}nigstuhl 17, 69117, Heidelberg, Germany \\
\altaffilmark{12}Space Science Institute, 4750 Walnut St., Suite 205, Boulder, CO, 80301, USA \\
\altaffilmark{13}Department of Physics, University of Cincinnati, Cincinnati, OH, 45221-0011, USA \\
\altaffilmark{14}Department of Physics and Astronomy, College of Charleston, 66 George St., Charleston, SC, 29424, USA \\
\altaffilmark{15}The Center for the Promotion of Integrated Sciences, The Graduate University for Advance Studies, Shonan International Village Hayama-cho, Miura-gun, Kanagawa, 240-0115, Japan \\
\altaffilmark{16}Institute for Astronomy, ETH Zurich, Wolfgang-Pauli-Strasse 27, 8093, Zurich, Switzerland \\
\altaffilmark{17}Laboratoire Hippolyte Fizeau, UMR6525, Universite de Nice Sophia-Antipolis, 28, avenue Valrose, 06108, Nice Cedex 02, France \\
\altaffilmark{18}Department of Astrophysical Sciences, Princeton University, Peyton Hall, Ivy Lane, Princeton, NJ, 08544, USA \\
\altaffilmark{19}12 Universit$\ddot{\rm a}$ts-Sternwarte M$\ddot{\rm u}$nchen, Ludwig-Maximilians-Universit $\ddot{\rm a}$t, Scheinerstr. 1, 81679 M$\ddot{\rm u}$nchen, Germany \\
\altaffilmark{20}Institute for Astronomy, University of Hawaii, 640 North A'ohoku Place, Hilo, HI, 96720, USA \\
\altaffilmark{21}Department of Astronomy, Kyoto University, Kita-shirakawa-Oiwake-cho, Sakyo-ku, Kyoto, 606-8502, Japan \\
\altaffilmark{22}Exoplanets and Stellar Astrophysics Laboratory, Code 667, Goddard Space Flight Center, Greenbelt, MD, 20771, USA \\
\altaffilmark{23}Hiroshima University, 1-3-2, Kagamiyama, Higashi-Hiroshima, Hiroshima, 739-8511, Japan \\
\altaffilmark{24}Space Telescope Science Institute, 3700 San Martin Drive, Baltimore, MD, 21218, USA \\
\altaffilmark{25}Center for Astrophysical Sciences, Johns Hopkins University, Baltimore, MD, 21218, USA \\
\altaffilmark{26}Jet Propulsion Laboratory, California Institute of Technology, 4800, Oak Grove Drive, Pasadena, CA, 91109, USA \\
\altaffilmark{27}Department of Astronomical Science, School of Physical Sciences, Graduate University for Advanced Studies (SOKENDAI), Mitaka, Tokyo, 181-8588, Japan \\
\altaffilmark{28}Kavli Institute for the Physics and Mathematics of the Universe, The University of Tokyo, 5-1-1, Kashiwanoha, Kashiwa, Chiba, 227-8568, Japan \\
\altaffilmark{29}Department of Cosmosciences, Hokkaido University, Kita-ku, Sapporo, 060-0810, Japan \\
\altaffilmark{30}Astronomical Institute, Tohoku University, Aoba-ku, Sendai, Miyagi, 980-8578, Japan \\
\altaffilmark{31}RESCEU, University of Tokyo, Hongo, 7-3-1 Bunkyo-ku, Tokyo, 113-0033, Japan \\
}




\begin{abstract}
We present a new Subaru/HiCIAO high-contrast \textit{H}-band polarized intensity (\textit{PI}) image of a nearby transitional disk associated with TW Hydrae. The scattered light from the disk was detected from 0\farcs2 to 1\farcs5 (11 -- 81 AU) and the \textit{PI} image shows a clear axisymmetric depression in polarized intensity at $\sim$ 0\farcs4 ($\sim$ 20 AU) from the central star, similar to the $\sim$ 80 AU gap previously reported from HST images. Azimuthal polarized intensity profile also shows the disk beyond 0\farcs2 is almost axisymmetric. We discuss two possible scenarios explaining the origin of the polarized intensity depression: 1) a gap structure may exist at $\sim$ 20 AU from the central star because of shallow slope seen in the polarized intensity profile, and 2) grain growth may be occurring in the inner region of the disk. Multi-band observations at NIR and millimeter/sub-millimeter wavelengths play a complementary role in investigating dust opacity and may help reveal the origin of the gap more precisely.
\end{abstract}






\section{INTRODUCTION}

The structures of young circumstellar disks can reveal key insights into the formation and evolution of planetary systems. Complex structures such as gaps, holes, spiral arms, or asymmetric structures have been imaged in a number of protoplanetary disks at near-infrared (NIR) and millimeter/sub-millimeter wavelengths \citep{hashimoto11,hashimoto12,andrews11,muto12,isella13,grady13,fukagawa13,mayama12,currie14} and they are predicted in theoretical simulations throughout the literature: a planet embedded in a viscous disk \citep{muto10,zhu11}, photoevaporation \citep{clarke01,Owen11a,Owen11b}, and grain growth \citep{birnstiel12b}.  

The 7--10 Myr old \citep{ducourant14} pre-main sequence M2 \citep{debes13} star TW Hydrae (hereafter TW Hya) shows particularly strong evidence for a disk making a \textit{transition} from a pre-planet building morphology to a post-planet building structure \citep[e.g.][]{strom89}. TW Hya, with 0.55 $\pm$ 0.15 $M_\sun$ in stellar mass \citep{debes13} and more than 0.05 $M_{\sun}$ in disk mass \citep{bergin13}, is an ideal example of transitional disks in that it is almost face-on \citep[${\rm i}$ = 7\arcdeg;][]{qi04} and because of its close distance of 54 $\pm$ 6 pc (Hipparcos; \citealt{vanleeuwen07}), is particularly accommodating to investigations of its geometrical structure with high spatial resolution and sensitivity. Previous studies at optical, near-IR, and mm/sub-mm wavelengths show the radial surface brightness extending to $\sim$ 150 -- 230 AU \citep{krist00,apai04,roberge05,weinberger02,qi04,qi06,andrews12}. High contrast direct imaging with the Hubble Space Telescope (HST) NICMOS coronagraph reveal a disk gap at 80 AU, indicating the evolution of dust composition or the presence of an embedded protoplanetary system \citep{debes13}. The inner hole at $\sim$ 4 AU is observationally predicted from a flux deficit at $\sim$ 2 -- 20 $\micron$ in the spectral energy distribution (SED), suggesting a deficit of material left around the central star. Excess at $\sim$ 20 -- 60 $\micron$ in the SED is probably due to emission from the truncated inner edge of the outer disk \citep{calvet02}. The existence of the inner disk is also supported by interferometric observations and visibility model fitting of the 7 mm continuum \citep{calvet02,hughes07} and a recent modeling concludes that the disk does not have a sharp edge at 4 AU \citep{menu14}.

In this Letter, we present a new $H$-band polarized intensity (\textit{PI}) image of TW Hya and the polarized intensity distribution ranging from 0\farcs2 to 1\farcs5 (corresponding to $\sim$ 11 -- 81 AU) of the disk, including the region closer to the central star than the region previously observed by HST. A polarized intensity depression at $\sim$ 0\farcs4 ($\sim$ 20 AU), probably caused by a gap in the disk, was detected. Through investigating observed polarized intensity profile, we discuss the disk morphological structure and the origin of the depression.  

\section{OBSERVATIONS AND DATA REDUCTION}
We carried out $H$-band ($1.6 \mu $m) linear polarimetric observations of TW Hya using the High Contrast Instrument for the Subaru Next Generation Adaptive Optics \citep[HiCIAO;][]{tamura06} with a dual-beam polarimeter at the Subaru 8.2 m telescope on March 24th 2011 and January 3rd 2013. These observations were part of the Strategic Explorations of Exoplanets and Disks with Subaru \citep[SEEDS;][]{tamura09} project. Both of the observations were conducted in the same instrument setting. The polarimetric observation mode acquires \textit{o}-ray and \textit{e}-ray simultaneously, and images a field of view 10\arcsec $\times$ 20\arcsec with a pixel scale of 9.50 mas/pixel.  

The exposures were performed at four position angles (P.A.s) of the half-wave plate, with a sequence of P.A. = 0\arcdeg, 45\arcdeg, 22.5\arcdeg, and 67.5\arcdeg, \ to measure the Stokes parameters. The integration time per wave plate position was set by 10 -- 30 s during observation to optimize instrument according to on-site condition. Three coadd modes were used to save the total overhead time. The adaptive optics system \citep[AO188;][]{hayano08} provided mostly stable stellar point-spread function (PSF) with a FWHM of 0\farcs15 in the \textit{H}-band. The adaptive optics corrections in both observation dates functioned well and the qualities of resultant images of both observations were almost the same. We strictly selected only good quality of data from among of them for producing the best image. Low quality images were removed prior to production of the final image, resulting in 13 good sets with a total integration time, for the \textit{PI} image, of 390 s. The data sets were taken with Angular Differential Imaging (ADI) mode with a net field rotation of 12 degrees. 

All unflagged data were corrected for field rotation and integrated. The standard imaging method \citep{hinkley09} was applied to the polarimetric data, including corrections for distortion and flat-field, using the Image Reduction and Analysis Facility (IRAF\footnote{IRAF is distributed by National Optical Astronomy Observatory, which is operated by the Association of Universities for Research in Astronomy, Inc., under cooperative agreement with the National Science Foundation.}) as in previous SEEDS publications \citep[e.g.][]{hashimoto11,muto12}.
Correction for instrumental polarization of the Nasmyth HiCIAO instrument was applied with the Mueller-matrices technique, that quantizes polarization characteristics \citep{joos08}. \textit{PI} was calculated from $\sqrt{Q^2+U^2}$, where $Q$ and $U$ are orthogonal linear polarizations, and the final polarized intensity profile was obtained after applying 10 $\times$ 10 pixels smoothing to the entire disk. 

\section{RESULTS}
\label{result}
\begin{figure}[h]
\begin{center}
\includegraphics[scale=0.40]{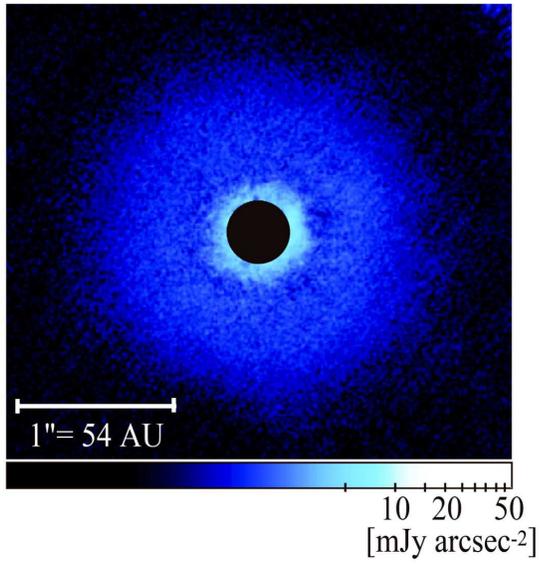}
\caption{\textit{H}-band \textit{PI} image of TW Hya in the ADI mode. North is up and east is to the left.
The dark filled circle at the center indicates a software mask with $r$ = 0\farcs2. (A color version of this figure is available in the online journal.)}
\label{fig:fig1}
\end{center}
\end{figure}

The \textit{PI} image in $H$-band of the transitional disk around TW Hya is presented in Figure \ref{fig:fig1}. There is no discernible misalignment in the polarization vector beyond $r = $ 0\farcs2 (which will be the subject of later studies in this series), and to be conservative, polarization measurements at radii more distant than 0\farcs2 from the central star are regarded as more accurate.

\begin{figure}[t]
\begin{center}
\includegraphics[scale=0.60]{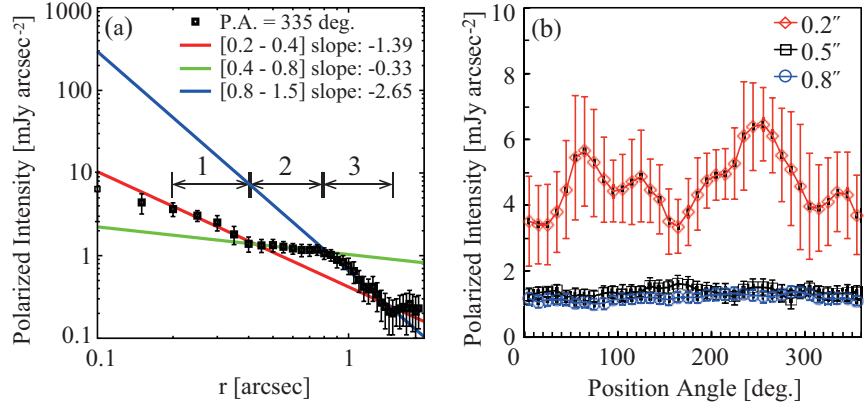}
\caption{Polarized intensity profiles of the TW Hya disk at $H$ band (left), and azimuthal brightness profile images at $r$ = 0\farcs2 ($\sim$ 11 AU), 0\farcs5 ($\sim$ 27 AU), and 0\farcs8 ($\sim$ 43 AU), respectively (right). The error bars shows 1 $\sigma$. The numbers 1--3 in the left panel denote the three radial zones for which a separate polarized intensity slope is fit.}
\label{fig:fig2}
\end{center}
\end{figure}

Shown in the \textit{PI} image is a ring-like polarized intensity depression at $r$ $\sim$ 0\farcs4. The observed polarized intensity profile shown in Figure \ref{fig:fig2}(a), revealing a stair-like decline in polarized intensity, with increasing distance from the central star. We define three major step transitions as: Zone 1 (0\farcs2 $\leq$ $r$ $\leq$ 0\farcs4), Zone 2 (0\farcs4 $\leq$ $r$ $\leq$ 0\farcs8), and Zone 3 (0\farcs8 $\leq$ $r$ $\leq$ 1\farcs5), and the polarized intensity profile is fitted at each zone with a simple power law: $\propto$ $r^{-\gamma}$. The fitting results in zones 1, 2, and 3 are shown in red, green, and blue, with slopes at each zone of -1.39, -0.33, and -2.65, respectively. 
The significant change in slope at $\sim$ 0\farcs4 is interpreted as a transition of some physical property in all azimuthal directions and is the subject of the remainder of this Letter. The change in slope around 0\farcs8 has already been discussed in \cite{krist00} and \cite{apai04}. 

Figure \ref{fig:fig2}(b) displays an azimuthal brightness profile at 0\farcs2, 0\farcs5, and 0\farcs8 in radius. This figure shows the disk is almost axisymmetric, with constant polarized intensity in all azimuthal directions, although rms noise level increases at $r =$ 0\farcs2.

\section{DISCUSSION}
One of interesting results in the observation is that the polarized intensity profile has a stair-like shape. As shown in Figure \ref{fig:fig2}a, in Zone 1, the innermost part of the disk, the profile shows a shallow slope of approximately $-1.4$, where the slope is expected to be $-2$ in the case of a flared disk, or steeper in the case of a flat geometry \citep{whitney92,inoue08}. In Zone 2, the profile appears almost flat, perhaps indicating that temperature increases with radius. In Zone 3, the polarized intensity profile has a steep value of approximately $-2.65$, which is darker than expected. The observed polarized intensity profile reflects a radial distribution of small dust particles and dust property. In this section, we discuss three possible explanations for the radial polarized intensity distribution observed in the disk of TW Hya. These include (a) an unusual radial distribution on the temperature, (b) the presence of a disk gap induced by the presence of a low-mass planet, and (c) a change in dust scale height due to grain growth.

\subsection{The Radial Temperature Gradient}
\label{temperature}

It is important to investigate the extent to which the shallow slopes observed in Zones 1 and 2 could depend on temperature. Generally, \citet{inoue08} shows the observed intensity of the scattered light ($I_{\rm obs}$) from a disk with no complex structures, such as gap or cavity, can be expressed by 
\begin{equation}
{I}_{\rm obs} \sim \frac{L_\star}{4\pi r^2}\beta \omega(\theta), 
\label{eq:intensity}
\end{equation}
where $L_\star$ is a stellar luminosity, $r$ is a distance from a central star, and $\omega(\theta)$ is a product of albedo and polarization (a phase function including polarization), $\theta$ is a scattering angle, and $\beta$ is grazing angle, and expressed by 
\begin{equation}
\beta \sim \frac{{\rm d}{H_s}}{{\rm d}r} - \frac{{H_{\rm s}}}{r} = r\frac{{\rm d}}{{\rm d}r}\left(\frac{{H_{\rm s}}}{r}\right),  
\label{eq:beta}
\end{equation}
where ${H_{\rm s}}$ is scattering surface. Assuming $\omega(\theta)$ is constant in Equation (\ref{eq:intensity}) and considering $I_{\rm obs} \propto r^{-k}$, we obtain 
\begin{equation}
\frac{{\rm d}}{{\rm d}r}\left(\frac{{H_{\rm s}}}{r}\right) \propto r^{1-k}. 
\label{eq:Hs/r}
\end{equation}
Thus, ${H_{\rm s}}$ becomes 
\begin{equation}
{H_{\rm s}}\propto r^{3-k}. 
\label{eq:Hs/r2}
\end{equation}

According to customary practice and precedents in the past works, we assume that ${H_{\rm s}}$ is linearly related to ${H_{\rm g}}$ by some constant factor A in the case of smoothly extended disk with no gap \citep{kenyon87,chiang97}, and the relation can be described by
\begin{equation}
{H_{\rm s}} \sim {\rm A}\frac{c_s}{\Omega_k} = {\rm A}{H_{\rm g}}, 
\label{eq:scat_surface2}
\end{equation} 
where $c_s$ is the speed of sound, $\Omega_k$ is Keplerian angular velocity, and $H_{\rm g}$ is a gas pressure scale height, respectively. Considering radial temperature distribution, $T(r)$, has a form of $T(r)$ $\propto$ $r^{-q}$, where $q$ is a power index, $c_s \propto r^{-q/2}$, and $\Omega_k \propto r^{-3/2}$, we obtain  
\begin{equation}
{H_{\rm s}} \propto r^{(3-q)/2}.  
\label{eq:scat_surface3}
\end{equation}

For Zone 1, with radial polarized intensity dropping as $r^{-1.39}$, Equation (\ref{eq:Hs/r2}) implies $H_{\rm s}$ proportional to $r^{1.61}$ and $q = - 0.22$. For Zone 2, with radial polarized intensity dropping as $r^{-0.33}$, $H_{\rm s}$ is proportional to $r^{2.67}$ and $q = - 2.34$. These indicate the temperature increases with increasing $r$, which is difficult to explain by considering only stellar radiation or flared structure of the disk. One of the reasons for the unexpected result is that the gas scale height is not simply related to temperature. Such a result might be able to be explained when a gap exists because the gap generates a sudden change of surface density. Dynamical phenomena such as disk wind or turbulence could be cited as another possible cause, because the disk surface can be lifted by them, resulting in a flat polarized intensity profile and no longer in hydrostatic equilibrium. If this is the case, wind mechanism such as photoevaopration or magneto-rotational instability that suppress the inner region of the disk due to dead zone might occur.

\subsection{Planet-Induced Disk Surface Structure}
\label{surface}

In this section, we consider the possibility that the changing slope of the observed polarized intensity distribution may signal the presence of a planet within the disk. The slopes at each zone can be explained by a disk surface geometry described by \citet{jang12,jang13}. In their works, the  disk illumination and shadowing are self-consistently calculated considering several disk orientations, using a static $\alpha$--disk radiative transfer model when a gap exists in the disk. Figure 2 in \citet{jang12} shows the surface brightness becomes brighter at the gap wall directly irradiated by a central star, and darker by shadowing and cooling behind the outer gap wall at the edge of gap. If this is the case, Zones 1, 2 and 3 probably correspond to a gap, an outer gap wall, and a shadowed region behind the outer gap wall, respectively. They also made radiative transfer calculations at several wavelengths and theoretically derived a radial surface brightness profile of a disk when a gap is formed by a 70 and 200 earth mass planet. The radial surface brightness profile at a wavelength of 1 $\micron$ shown in Figure 6 in \citet{jang12} increases just beyond the gap created by a planet, and its slope depends on planet mass, flatter in the case of small mass planet. Therefore, if the origin of the polarized intensity depression attributes to a planet, planet mass seems to be small since our observed polarized intensity profile shows a flat slope at the gap region. 

It is beyond the scope of this paper to investigate the detailed effects of planet parameters on the structure of the inner disk of TW Hya, but an extensive search of parameter-space (planet mass and orbital characteristics) in the future might lead to a suitable model for the observed brightness distribution.


\subsection{Possibility of Grain Growth}
\label{grain_growth}
The picture of the disk structure presented in Section \ref{temperature} and \ref{surface} does not necessarily mean that the gas disk has a gap at $r \sim$ 20 AU. Dust grains are also believed to settle to the mid-plane as their size increases. When the disk gas has a Gaussian distribution in the vertical direction with $H_{\rm g}$, dust grains also have a Gaussian distribution in the vertical direction, with a dust scale height $H_{\rm d}$ that is determined by the balance between the diffusion by the gas and the coupling efficiency to the gas \citep{dubrulle95}. Using the Stokes number St, which represents the coupling efficiency of grains to gas and is defined as the stopping time normalized by the orbital timescale, and the diffusion coefficient $\alpha_{\rm D}$, which represents the diffusion efficiency of the gas in the vertical direction, the dust scale height can be written as 

\begin{equation}
H_{\rm d}=H_{\rm g}\max
 \Bigl[1, \left(\frac{\alpha_{\rm D}}{\rm St}\right)^{0.5}\Bigr],
\label{eq:dustscaleheight}
\end{equation}
when St $\ll$ 1 \citep[see][and references therein]{brauer08}.
As a first rough step to quickly quest to see a possibility whether grain growth occurs or not, we introduce the parameter $s$ where the radial grain size distribution has a form of $a(r) \propto r^{-s}$. By using $\tilde{p}$ in the form of $\Sigma_{\rm g}\propto r^{-\tilde{p}}$ and applying turbulent viscosity $\alpha$ model \citep{shakura73} and the Epstein law, then the Stokes number has a simple form of 
\begin{equation}
 {\rm St} \sim \frac{\rho_{\rm d} a(r)}{\Sigma_{\rm g}} \propto r^{\tilde{p}-s},
\end{equation}
where $\rho_{\rm d}$ is the density of a dust particle and assumed to be a constant throughout the disk. Substituting this into Equation (\ref{eq:dustscaleheight}), we obtain
\begin{equation}
 H_{\rm d} \propto H_{\rm g} r^{(s-\tilde{p})/2}.
\end{equation}
Since $H_{\rm g} \propto r^{(3-q)/2}$, we finally obtain the radial dependence of the dust scale height as 
\begin{equation}
 H_{\rm d} \propto r^{(3+s-\tilde{p}-q)/2}.
  \label{eq:final}
\end{equation}

Our observation suggests that the dust scale height has 
a radial dependence of $H_{\rm d} \propto r^{1.61}$ in Zone 1 and $H_{\rm d} \propto r^{2.67}$ in Zone 2, by utilizing Equation (\ref{eq:Hs/r2}) and assumption that $H_s$ is linearly related to $H_g$ described by Equation (\ref{eq:scat_surface2}). Note that $H_{\rm d}$ and $H_{\rm s}$ are not identical, however $H_{\rm s}$ must be about equal to $H_{\rm d}$ for the small dust grains, which are mainly traced by $H$-band scattering light in our observation. From Equation (\ref{eq:final}), we obtain $s=1.72$ in Zone 1 and $s=3.84$ in Zone 2 in the disk around TW Hya, assuming that $\tilde{p}=1$ and $q = 0.5$ \citep{hughes08}. This suggests that the dust has a larger size in the inner part of the disk, and the size varies by at least a factor of 10 between 10 AU ($\sim 0\farcs2$) and 40 AU ($\sim 0\farcs8$), where the depression and its vicinity are covered.

It should be noted that since smaller grains have a larger scattering efficiency, the scattering plane has to represent the dust scale height of the smallest grains at the radial position. The $PI$ image suggests the possible existence of a small-grain depleted region at $r< 0\farcs4$, with the depleted range of width more than 10 AU wide. This may suggest that the larger grains are present at inner radii, but there is no sharp transition of the grain size distribution. We note that gas and large grains can be present in the inner region of the disk. Indeed, mm-/submm-observations give no evidence to date of the depletion of gas or dust continuum emission. Furthermore, the dust sedimentation near the disk mid-plane results in flat polarization intensity profile, that is it further promotes a flared structure in $H_{\rm d}$ distribution.

Estimates of the amount of 10 $\micron$ -- 1 mm sized dust in the disk mid-plane, and up to $\sim$ 60 AU, are limited by SMA and VLA data \citep[see Fig. 9,][]{menu14}. Our NIR observations trace dust at the disk surface between 10 -- 80 AU, and reflect mainly smaller grains 0.1 -- 10 $\micron$ in size. The observations are complementary to each other, revealing radially and vertically distributed dust at different size scales. Note that the absorption efficiency also depends on grain size, and thus influences on the vertical profile via temperature distribution. However, the absorption efficiency does not significantly change with dust grain sizes between 0.1 -- 10 $\micron$ in the case of $H$-band wavelength \citep{miyake93}.

The variation of fundamental dust properties may also affect the observed polarized intensity distribution. For instance, $\omega(\theta)$ in Equation (\ref{eq:intensity}) may not be constant. \citet{takami13} explored the scattering properties for different grain size, and calculate polarization degree and polarization intensity normalized by the incident stellar light as a function of scattering angle. They show the polarization degree decreases with increasing grain size when $2\pi a<\lambda$, where $a$ is a radius of a dust grain and $\lambda$ is wavelength of emission, respectively, resulting in a depolarization of the emission \citep[see also][]{fischer94}. On the other hand, when $2\pi a>\lambda$, polarization and polarized intensity decreases less significantly even though $a$ becomes larger. Since the dust grain size directly affects the polarized intensity, a variable $\omega(\theta)$ is consistent with the effect of grain growth.

The minimum grain size depends strongly on the fragmentation efficiency of dust-dust collisions in the disk. If fragmentation of dust is efficient, the scattering surface of the disk should be dominated by the small fragments. However, our observational analysis suggests that fragmentation is not so efficient, particularly in the inner part of the disk. One possible reason for the inefficiency of fragmentation is the icy aggregates have a larger critical velocity before complete disruption, than do rocky aggregates \citet{Wada09, Wada13}, although the critical velocity of fragmentation is still under debate. In our study, it is reasonable to assume that the grains are covered by ice, since the locations we have observed are expected to be outside the water snow line.  

Another possibility is that fragmentation is effective, but the tiny fragments are swept up by larger grains. \citet{birnstiel12b} investigated the size distribution as a result of the balance between fragmentation and sweep-up. To sweep up tiny grains, they suggest the diffusion coefficient $\alpha$ should be as low as $\alpha_{\rm D}=10^{-4}$. In our case, the modeling therefore suggests the diffusion is generally inefficient throughout the disk.

In tandem with NIR high-resolution observations, high-resolution and multi-band observations at mm-wavelength will aid an investigation into the grain opacity index $\beta$. Such data will help us to clarify whether the grain growth is occurring at inner radii and understand the effects of any planet formation on dust character within the gap.



This work is partially supported by KAKENHI 22000005(M.T.), KAKENHI 23103004, 23740151 (M.F.), NSF AST 1008440 (C.A.G.), NSF AST 1009314 (J.P.W.), NASA NNX09AC73G (C.A.G. and M.L.S.). C.A.G. was also supported under NSF AST 1008440 and through the NASA Origins of Solar Systems program on NNG13PB64P. J. K. is supported by Grant-in-Aid for JSPS Fellows (26$\cdot$04023). T.M. is supported by JSPS KAKENHI Grant Numbers 26800106, 23103004, 26400224. 
J. C. is supported by a grant from the U.S. National Science Foundation under Award No. 1009203. M. T. is supported from Ministry of Science and Technology (MoST) of Taiwan (Grant No. 103-2112-M-001-029).

\acknowledgments

\clearpage

\clearpage

\end{document}